\documentclass[aps,prd,preprint,letterpaper,nofootinbib,floatfix,amsmath,amssymb]{revtex4}

\usepackage{subfig}
\usepackage{graphicx}

\newcommand{\unit}[2][]{\text{#1}\,\text{#2}}
\newcommand{\ds}{ }
\newcommand{\diff}[1]{\text{d}#1} 
\newcommand{\abs}[1]{\left|#1\right|} 
\newcommand{\bra}[1]{\left\langle #1\right|}
\newcommand{\ket}[1]{\left|#1\right\rangle}
\newcommand{\braket}[2]{\left\langle#1|#2\right\rangle}
\newcommand{\bigO}[1]{\mathcal{O}(#1)} 
\newcommand{\ident}{\text{{\small $1$}}\hspace{-0.37em}1}
\newcommand{\Ex}[1]{\text{Ex}\left(#1\right)} 
\newcommand{\Exx}[1]{\text{Ex}(#1)} 

\newcommand{\vecN}[1]{\hat{#1}}
\newcommand{\tot}[1]{{#1}_\text{tot}}
\newcommand{\Etot}{\tot{E}}
\newcommand{\bs}{\boldsymbol}
\newcommand{\fv}{\tilde{f}}

\newcommand{\trk}{\text{trk}}
\newcommand{\twr}{\text{twr}}
\newcommand{\coeff}[1]{{}\bar{#1}}

\begin{document}

\title{Harnessing the global correlations of the QCD power spectrum}

\author{Keith~Pedersen}
\affiliation{Department of Physics, Illinois Institute of Technology, Chicago, Illinois 60616-3793, USA}

\author{Mithila~Mangedarage}
\affiliation{Department of Physics, Illinois Institute of Technology, Chicago, Illinois 60616-3793, USA}

\author{Zack~Sullivan}
\email{Zack.Sullivan@IIT.edu}
\affiliation{Department of Physics, Illinois Institute of Technology, Chicago, Illinois 60616-3793, USA}

\date{December 10, 2023}

\begin{abstract}
As multiplicity increases at the CERN Large Hadron Collider, an
opportunity arises to explore the information contained in the full
QCD power spectrum on an event-by-event basis.  This paper lays the
foundations for a framework to encode and extract the information
contained in finite sampling of a QCD event.
\end{abstract}

\maketitle

\section{Introduction}

While study of the QCD radiation spectrum at colliders has been
dominated in recent decades by the study of jets
\cite{Ellis:1980wv,Ellis:1993tq,Cacciari:2008gp} and jet substructure
\cite{Larkoski:2017jix,Asquith:2018igt}, modern colliders are faced
with phenomena that motivate a return to a more continuous, global
approach to QCD events. The long-distance, same-side ridge correlation
seen in lead ion collisions \emph{also} appears in high-multiplicity
proton
collisions~\cite{Khachatryan:2010gv,TheATLAScollaboration:2015uzr},
and a satisfactory explanation will likely require a better
understanding of the inter-particle correlations.  In short-distance
physics, the reconstruction of two \unit[40]{GeV} jets from a $W^\pm$
becomes a challenging affair when the final state is bathed in
hundreds of pileup vertices \cite{Calkins:2013ega}.  This paper
revisits the global approach of the QCD power spectrum, identifies and
solves technical challenges with its use, and introduces a new
framework in which to study the power spectrum on an event-by-event
basis.

Fox and Wolfram created a framework to study the angular power
spectrum of QCD in their seminal papers~\cite{Fox:1978vu,Fox:1978vw},
which began the study of average event shapes in $e^+e^-$ colliders
\cite{Ellis:1980nc,Ellis:1980wv,Vermaseren:1980qz}. Despite a long
history \cite{Ali:2010tw} of studies of the energy-energy correlations
(EEC) in the QCD power spectrum
\cite{Basham:1978bw,Fox:1980tz,Ali:1982ub}, previous papers have not
fully addressed a significant problem: QCD radiation can only be
observed through a \emph{finite} sampling of \emph{discrete}
particles. Discrete samples have a limited sampling frequency, and
inaccessible frequencies must be removed to prevent aliasing in an
analysis.  This effective band limit is normally set by the
limitations of the measuring device or analysis (e.g., the windowing
function in CMB analyses~\cite{1995ApJ...443....6W}), but QCD events
are fundamentally different.  The smallest angular scale for which
there is meaningful information is determined by the nature of the
event itself. In this paper we will see that the primary factors
governing an event's angular resolution are the particle multiplicity
and general topology.

In Section~\ref{sec:power-spectrum} we solve two important challenges
to constructing a robust representation of the QCD power spectrum in
an event: (i)~a finite sample has an intrinsically limited angular
resolution, and (ii)~depending on the method of reconstruction,
detector objects have different resolutions (e.g., a charged track has
much better angular resolution than a calorimeter tower).  Both
challenges are solved in Sec.\ \ref{sec:shape-functions} by
introducing ``shape functions'' that impose a spatial band limit to
remove meaningless small-angle correlations.  Not only are shape
functions necessary for a complete decomposition of the QCD power
spectrum, they also guarantee collinear safety~\cite{Fox:1980tz}.  Our
framework should be useful for \emph{any} angular-correlation analysis
of particle-physics events, not just angular power spectra.  In
Section~\ref{sec:conclusions} we suggest future applications of the
power spectrum to better understand QCD, including tests of jet
substructure, hadronization models, and phenomenology at the CERN
Large Hadron Collider (LHC).  Derivations of some results are
collected in the Appendices.

\section{The angular power spectrum of QCD}\label{sec:power-spectrum}

The angular power spectrum encodes the amount of energy correlated at
various angular separations, and provides a natural framework for
sorting QCD radiation into hard and soft components.  In this section,
we begin by reviewing the definition first introduced by Fox and
Wolfram \cite{Fox:1978vu,Fox:1978vw}, and use predictions for two and
three jet-like events to uncover how finite particle number introduces
an effective angular resolution for an event and other artifacts into
the spectrum.

In the center-of-momentum (CM) frame of a QCD event with sufficiently
large interaction scale~$Q$, the final-state particles which are
measurable and relevant are (i) effectively massless and (ii) travel
radially outward from the primary vertex.\footnote {\nobreak After
  correcting for in-flight decays and the deflection of charged
  particles in the detector's magnetic field.  } Thus, all usable
information about QCD radiation is encoded in the event's
angularly-correlated energy density
\begin{equation}\label{eq:E-density}
	E(\vecN{r}) 
		= \sum_{i=1}^N E_i\ds\,
		\frac{\delta^2(\vecN{r} - \vecN{p}_i\ds)}
		{\sin\theta_i\ds}
	\,,
\end{equation}
which portrays a set of $N$ massless particles with 
energy $E_i\ds$ and radial direction of travel~$\vecN{p}_i\ds$.
$E(\vecN{r})$ is directly proportional to 
$E_\text{tot}\equiv\sum_i E_i\ds = \int_\Omega \diff{\Omega}\,E(\vecN{r})$,
so to isolate the radiation pattern we can remove this 
scalar degree of freedom by defining the dimensionless ``event shape''
\begin{equation}\label{eq:event-shape}
	\rho(\vecN r) \equiv
		\frac{E(\vecN r)}{E_{\rm tot}}
	\,.
\end{equation}
Fox and Wolfram decomposed this shape into the complete, orthonormal basis of 
the spherical harmonics~$Y_\ell^m$
\begin{equation}\label{eq:Ylm-complete}
	\rho(\vecN{r}) =
		\sum_{\ell=0}^{\infty} \sum_{m=-\ell}^{\ell} \rho_\ell^m\, Y_\ell^{m}(\vecN{r})
	\,,
	\quad\text{with}\quad
	\rho_\ell^m = \int_\Omega\diff{\Omega}\,{Y_\ell^m}^*(\vecN{r})\,\rho(\vecN{r})
	\,.
\end{equation}
Summing the coefficients of this $Y_\ell^m$ decomposition over $m$ reveals
how closely the event resembles radiation with an $\ell$-prong shape
--- this is the dimensionless ``power spectrum''
\begin{equation}\label{eq:H_l-continuous}
	H_\ell\ds\equiv
		\frac{4\pi}{2\ell+1}
		\sum_{m=-\ell}^{\ell}\abs{\rho_\ell^m}^2 
		= \int_{\Omega}\diff{\Omega}\int_{\Omega^\prime}\diff{\Omega^\prime}\,
		\rho(\vecN{r})\rho(\vecN{r}^\prime)
		P_\ell\ds(\vecN{r}\cdot\vecN{r}^\prime) \,,
\end{equation}
which introduces the Legendre polynomial $P_\ell\ds$ via the
$Y_\ell^m$ addition theorem.

The power spectrum exists on the unit interval ($0\le H_\ell\ds\le1$),
so $H_\ell\ds\approx1$ indicates the event exhibits a very
$\ell$-prong-like shape.  The power spectrum is invariant to
rotations, since SO(3) preserves the vector dot product (the only part
of Eq.~\ref{eq:H_l-continuous}'s integral which is not factorizable).
Thus, in the CM frame of a generic QCD event, the event's absolute
orientation does not matter; the power spectrum depends only on an
event's topology.  Na\"ively, one expects a very two-jet-like event
(one in which the event's energy is predominantly back-to-back) to
have a significant \hbox{2-prong} component, and therefore a large
value of $H_2\ds$.  The same expectation can be drawn for a 3-jet-like
event, whose three localized bundles of energy should create a large
value of~$H_3$.  Normalization requires that
$\int_\Omega\diff{\Omega}\,\rho(\vecN{r}) = 1 =
\sqrt{4\pi}\,\rho_0^0$, so $H_0\ds = 1$ for every event.

With massless partons, an all-orders prediction for QCD radiation in a
collider event would produce an infinite number of partons, and thus a
fully continuous $\rho(\vecN{r})$.  Not only is such a prediction 
intractable, the non-zero light-quark mass requires that QCD events must
contain a \emph{finite} number of particles at \emph{discrete}
locations.  The simplest distribution to describe this radiation is
the Fox-Wolfram (discrete) event shape,
\begin{equation}\label{eq:event-shape-delta}
	\rho(\vecN{r}) 
		= \sum_{i=1}^N f_i\ds\,
		\frac{\delta^2(\vecN{r} - \vecN{p}_i\ds)}
		{\sin\theta_i\ds}
	\,,
\end{equation}
which uses each particle's energy fraction
\begin{equation}
	f_i\ds \equiv \frac{E_i\ds}{\Etot\ds}
	\,.
\end{equation}
When calculating the power spectrum of the Fox-Wolfram shape,
the $\delta$-distributions describing each particle's spatial location 
collapse the integral to linear algebra equation
\begin{align}\label{eq:H_l-discrete}
	H_\ell\ds 
		& = 
		f_i\ds P_\ell\ds(\vecN{p}_i\ds\cdot\vecN{p}_j\ds) f_j\ds
		= \bra{f} P_\ell\ds\big(\ket{\vecN{p}} \cdot\bra{\vecN{p}}\big)\ket{f}
	\,.
\end{align}
It is crucial to distinguish this $H_l\ds$, the result of choosing the
Fox-Wolfram event shape, from the canonical definition of the power
spectrum (Eq.~\ref{eq:H_l-continuous}), which is agnostic to the
choice of~$\rho(\vecN{r})$.  By keeping particles discrete, the
Fox-Wolfram event shape attempts to measure the event with infinite
precision.  We will see the consequences of this in the next
subsection.

\subsection{The power spectrum of simple 3-jet events}

\begin{figure*}[htb]
\subfloat[\label{fig:2-jet-unsmeared}2-jet-like]{
\includegraphics[width=0.5\textwidth]{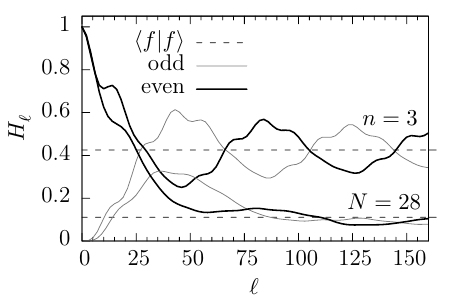}
\hspace*{-21ex}\raisebox{19.5ex}{\includegraphics[width=10ex]{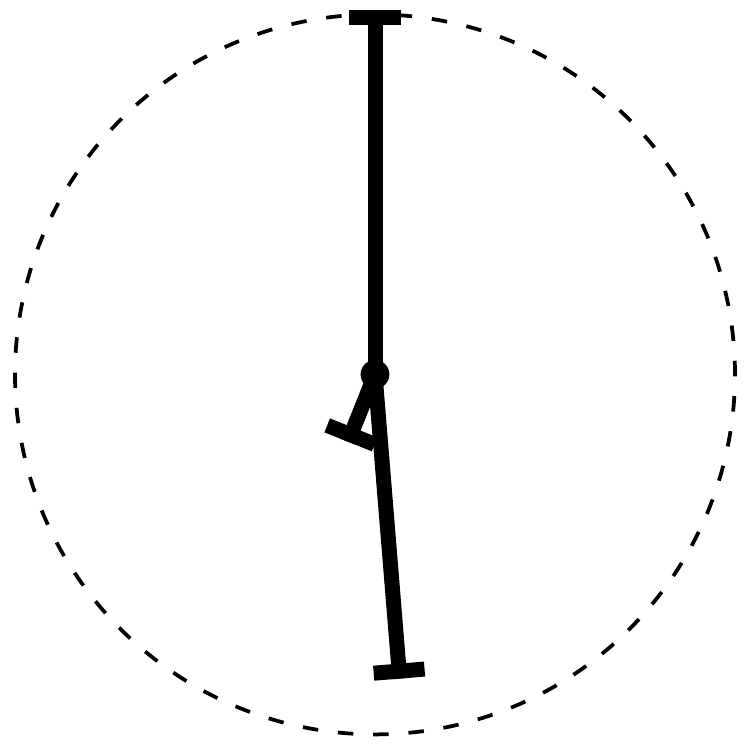}}\hspace{11ex}
}
\subfloat[\label{fig:3-jet-unsmeared}3-jet-like]{\includegraphics[width=0.5\textwidth]{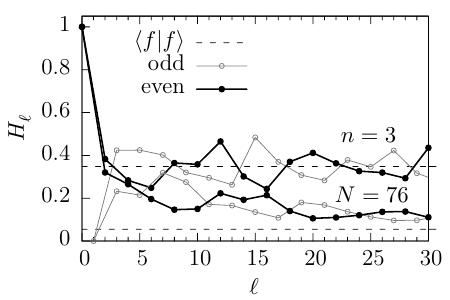}
\hspace*{-21ex}\raisebox{19.5ex}{\includegraphics[width=10ex]{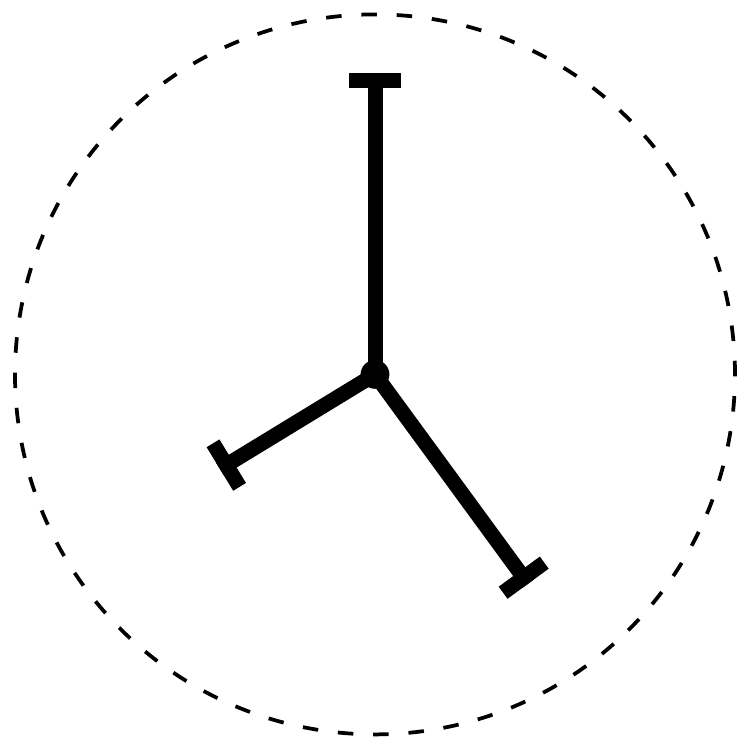}}\hspace{11ex}
}
\caption{The power spectra for two $e^+e^-\to q\bar{q}g$ events at $\sqrt{S}=\unit[250]{GeV}$ which are
(a)~2-jet-like and (b)~3-jet-like. Lines connect even and odd moments at integer $\ell$. 
Each figure shows $H_\ell\ds$ for (upper) $n=3$ initiating partons and
(lower) $N=\bigO{50}$ measurable particles (after showering and hadronization).
A dotted line shows the value of $\braket{f}{f}$ for each series, 
and the inset depicts the energy and orientation of the original partons.}
\label{fig:Hl-intro}
\end{figure*}

Before we discuss QCD power spectra, it is useful to conceptualize how
the energy is distributed by studying $H_\ell\ds$ for a few simple
events.  We begin by simulating $e^+e^-\to q\bar{q}g$ at
${\sqrt{S}=\unit[250]{GeV}}$ using MadGraph~5~\cite{Alwall:2014hca} at
tree-level.  These events are then showered and hadronized using
Pythia~8~\cite{Sjostrand:2006za,Sjostrand:2007gs}.
Figure~\ref{fig:Hl-intro} shows the power spectra for two such events,
using the Fox-Wolfram event shape of Eq.~\ref{eq:event-shape-delta}.
Note that $H_\ell\ds$ for the trivial QCD event --- two back-to-back
$\delta$-distributions in their CM frame --- is
\begin{equation}\label{eq:2-parton-Hl}
	H_\ell\ds = 
	\begin{cases}
	1 & \ell=\text{even}\\
	0 & \ell=\text{odd}
	\end{cases}
	\,.
\end{equation}

The power spectrum for a very two-jet like event is shown in
Fig.~\ref{fig:2-jet-unsmeared}, where we connect even (black lines)
and odd (gray lines) moments to aid the eye.  To highlight the
importance of jet formation on the QCD radiation spectrum, we show
$H_\ell\ds$ calculated for the ${n=3}$ original partons (upper pair of
lines), and the ${N=28}$ particles (hadrons and leptons and photons)
measurable by a detector (lower pair of lines).  For $\ell<10$, the
even moments are large and odd moments small --- matching the
prediction of Eq.~\ref{eq:2-parton-Hl} --- with the $H_\ell\ds$ for
the measurable particles closely following that of their originating
partons. This close correspondence between extensive jets and
infinitesimal partons derives from the angular scale $\xi$ of each
$H_\ell\ds$:
\begin{equation}\label{eq:Hl-angular-scale}
	\xi=\frac{2\pi}{\ell}
	\,.
\end{equation}
Low-$\ell$ moments have a coarse angular scale, and are not terribly
sensitive to the jet shape.  However, as $\ell$ increases, $H_\ell\ds$
begins to detect the jets' spatial extent, and the two series diverge.
Each power spectrum eventually stabilizes to
$H_\ell\ds\sim\braket{f}{f}$ (shown with a dotted line).

The power spectra of the 3-jet-like event in
Fig.~\ref{fig:3-jet-unsmeared} is very different from the 2-jet-like
event.  As expected, an event with three distinct jets has a
significant 3-prong power, since $H_3\ds\approx H_2\ds\approx1/3$.
Jet structure has a stronger influence over $H_\ell\ds$, since the
measurable and 3-parton spectra diverge at a much lower $\ell$.
However, the most critical feature is the flattening of
$H_\ell\ds$ in Fig.~\ref{fig:Hl-intro} to an asymptotic plateau at
$\braket{f}{f}$, which drives a divergence in the total power.  This
is a symptom of a serious flaw in $H_\ell\ds$; the
$\delta$-distributions in the series approximation of $\rho(\vecN{r})$
in the $Y_\ell^m$ basis (Eq.~\ref{eq:Ylm-complete}) is incomplete and
never converges.  A $Y_\ell^m$ decomposition is only complete when the
function is square-integrable (i.e., $\int_\Omega
\diff{\Omega}\,\rho^2(\vecN{r}) < \infty$), and $\delta$-distributions
are not.  We will see next that this problem is related to the limited
angular resolution of a finite sample, and develop a remedy.

\subsection{Angular resolution of a finite sample}

\begin{figure}[htb]
\subfloat[\label{fig:iso-trk}tracks]{\includegraphics[width=0.5\textwidth]{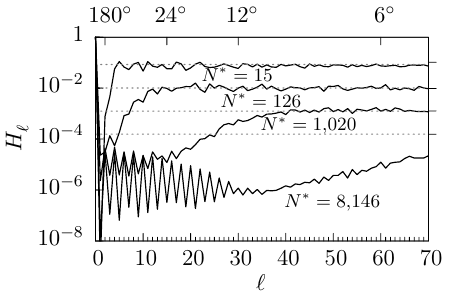}}
\subfloat[\label{fig:iso-twr}towers]{\includegraphics[width=0.5\textwidth]{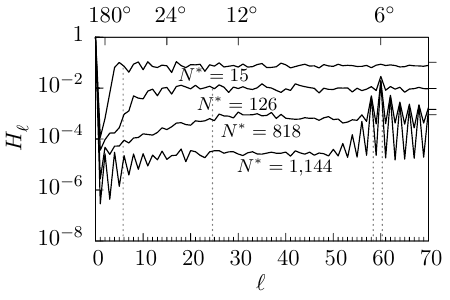}}
\caption{
The power spectrum of isotropic samples of size $N=\lbrace16, 128,
1024, 8192\rbrace$, seen with a pseudo-detector (using 1,144 towers of
$\Omega_\twr\ds \approx (6^\circ)^2$,
$\eta_{\max}^\trk=\eta_{\max}^\twr\approx3$, and $p_T^{\min}=0$).
Each series is labeled by the number of objects $N^*$ actually
detected, with (a)~the entire sample detected as charged \emph{tracks}
and (b)~the entire sample detected as neutral \emph{towers}.  A
horizontal tick on the right edge depicts each
sample's~$\braket{f}{f}$.
}
\label{fig:iso}
\end{figure}

In order to understand the power spectrum's asymptotic plateau, and
certain detector artifacts which we expect to see in measured power
spectra, we briefly examine a toy model.  Imagine a collision process
that scatters particle energy isotropically and homogeneously across
the sphere.  This model's trivial event shape $\rho(\vecN{r}) =
1/(4\pi)$ has a featureless power spectrum: $H_\ell\ds=0$ for
$\ell>0$.  Yet when we simulate this process, the power spectra are
not featureless --- primarily because each event has a \emph{finite}
number of particles (and is therefore inhomogeneous), but further
because detectors do not perfectly measure particle positions.  Two
irreducible detector effects create important angular artifacts:
(i)~the inactive beam hole, and (ii)~the reduced angular precision of
calorimeter towers.  To study these effects, we construct a nearly
truth-level pseudo-detector,\footnote {A calorimeter with perfect
  energy resolution, built from towers of solid
  angle~$\Omega_\twr\ds$, detects particles out to $\eta_{\max}^\twr$
  (the edge of the beam hole).  Since neutral particles are not
  tracked, each tower is treated as massless object with energy
  $E_\twr^0 = E_\twr\ds - \underset{\text{tracks}}{\sum_i}
  \abs{\vec{p}_i\ds}$, where the momentum from perfectly reconstructed
  massless charged particles is subtracted.}
through which we pass isotropic events built from $N$ generic
particles.  Each event is detected \emph{twice}; first by making every
particle charged (a track-only detection to study the beam hole
effect), then by making all particles neutral (a tower-only detection
to study the calorimeter effect).  Each track or tower becomes a
$\delta$-distribution in the Fox-Wolfram event shape
(Eq.~\ref{eq:event-shape-delta}), whose $H_\ell\ds$ is calculated via
Eq.~\ref{eq:H_l-discrete}.  The power spectra of four such events,
with widely varying particle multiplicity~$N$, are shown in
Fig.~\ref{fig:iso} (which now connect \emph{adjacent} $\ell$ with
lines).

Figure~\ref{fig:iso-trk}'s track-only power spectrum confirms the
isotropy of the samples; ${H_\ell\ds\approx0}$ for $\ell>0$ (note the
log-scale).  As particle multiplicity increases, the samples become
increasingly homogeneous, further diminishing the residual non-zero
power.  But the falling power uncovers an oscillation at low~$\ell$, a
symptom of the handful of particles which escape detection through the
beam holes.  Since the angular scale of each moment in $H_\ell\ds$ is
$\xi=2\pi/\ell$ (Eq.~\ref{eq:Hl-angular-scale}), the correlations
between two opposing holes emanate from their fundamental ${\ell=2}$
mode.\footnote
{While the $\ell=1$ ``dipole'' has two back-to-back lobes, they have
  opposite polarity.  Two back-to-back shapes with the \emph{same}
  polarity (e.g., opposing jets or beam holes) register at $\ell=2$.
}
This beam-hole artifact is fully evident only in the
high-multiplicity samples because one must cover the sphere quite
thoroughly before detecting two small holes.  Fortunately, the
$\bigO{10^{-5}}$ power of the beam-hole artifact is quite small, so it
is swamped by random sampling noise in low-multiplicity samples.

Another important feature in Fig.~\ref{fig:iso-trk} is the power
spectrum's flat asymptotic plateau We can explain this plateau by
splitting the power spectrum of the Fox-Wolfram event shape into
constant~\emph{self} and $\ell$-dependent \emph{inter-particle}
correlations:
\begin{equation}\label{eq:Hl-asym}
	H_\ell\ds = \underset{\text{self}}{\underbrace{\braket{f}{f}}} + 
	\underset{\text{inter-particle}}{\underbrace{\bra{f}
	\left(P_\ell\ds\big(\ket{\vecN{p}}\cdot\bra{\vecN{p}}\big) - \ident\right)\ket{f}}}
	\,.
\end{equation}
As $\ell$ increases, inter-particle correlations are analyzed at
increasingly smaller angular scales.  Eventually, one reaches
$\xi_{\min}\ds$: the angular scale of the smallest inter-particle
angles.  At smaller scales (higher $\ell$), $H_\ell\ds$ begins to
probe the empty space between nearest neighbors, which is dominated by
random sampling noise (i.e., the \emph{exact} location of each
\mbox{$\delta$-distribution} relative to all others); this causes the
inter-particle correlations to interfere destructively.  The remaining
``self'' term creates a featureless, asymptotic plateau at
${H_\ell\ds\sim\braket{f}{f}}$ (each sample's $\braket{f}{f}$ is shown
with a horizontal tick on the graph's right edge).

Since the asymptotic plateau is born in the transition from useful
information to noise, the plateau effectively defines a power
spectrum's \emph{angular resolution}.  The high-multiplicity samples
in Fig.~\ref{fig:iso-trk} have lower plateaus which begin at higher
$\ell$, so they are better able to discern the high-$\ell$ components
of the beam-hole artifact.  This relationship to multiplicity is made
more explicit in Appendix~\ref{sec:ff-N}, where we show that
$\braket{f}{f}$ has an expected value of
\begin{equation}\label{eq:plateau-height}
	\Ex{\braket{f}{f}}=\frac{1+a}{N} \,,
\end{equation}
where $a\ge0$, and depends on the probability distribution of the
particles' energy fraction.  The dotted lines in
Fig.~\ref{fig:iso-trk} show this prediction for $a\approx0.278$ (a
value determined in Ref.~\cite{Pedersen2018} for our isotropic
sampling); this prediction agrees quite well with each plateau's
actual height.  Unfortunately, Eq.~\ref{eq:plateau-height} does not
divulge the angular resolution, it merely shows that it is somehow
related to multiplicity.

To quantify a sample's angular resolution, we examine
Fig.~\ref{fig:iso-twr}, whose tower-only power spectrum induces an
artificial $\ell=60$ correlation (corresponding to the $6^\circ$
separation of nearest neighbor towers).  While this occurs in every
sample, its angular artifact is only visible in the high-multiplicity
samples where particles have activated multiple towers.  Ultimately,
the poor angular resolution of the low-multiplicity samples is driven
by their sparseness; in the $N^*=15$ sample of Fig.~\ref{fig:iso-twr},
the closest particles are $18^\circ$ apart --- too distant to uncover
a $6^\circ$ correlation.  This further demonstrates that $H_\ell\ds$
cannot resolve correlations much beyond the smallest inter-particle
angles.  Furthermore, multiplicity is not the only factor; 15
particles in two intensely collimated, back-to-back jets will have a
much finer angular resolution than 15 particles distributed
isotropically, since the collimated particles in the former will be
much closer together.

Given that the scale of the smallest inter-particle angles defines a
sample's angular resolution, we give $\xi_{\min}\ds$ a rigorous
definition by appealing to the $H_\ell\ds$ inter-particle term in
Eq.~\ref{eq:Hl-asym}.  First, the symmetric matrices of inter-particle
angles ${\xi_{ij}\ds\equiv\arccos(\vecN{p}_i\ds\cdot\vecN{p}_j\ds)}$
and correlation weights $w_{ij}\ds = f_i\ds f_j\ds$ are flattened into
vectors $\ket{\xi}$ and $\ket{w}$ (keeping only $j>i$, since the
lower-half of $\xi_{ij}\ds$ is redundant, and its diagonal is null).
We then sort $\ket{\xi}$ from smallest to largest, sequencing
$\ket{w}$ to this new order.  A legitimate inter-particle correlation
must stand out from the asymptotic plateau of self-correlations, so we
find the first $n$ angles whose collective weight is larger than
$\braket{f}{f}$ (i.e., $n$ is the first index where $\sum_{k=1}^n
2\,w_k\ds \ge \braket{f}{f}$).\footnote
{Since $\Ex{\braket{f}{f}}\propto N^{-1}$ and $\Ex{f}=N^{-1}$,
  $n=\bigO{N}$.  The weight $w_k\ds$ is doubled because $\xi_{ij}\ds$
  is symmetric.}
The angular resolution $\xi_{\min}\ds$ can then be
computed from the weighted geometric mean of these $n$ smallest
angles:
\begin{equation}\label{eq:angular-resolution}
	\xi_{\min}\ds = \exp\left(\frac{\sum_{k=1}^{n} w_k\ds \,\log\xi_k\ds}{\sum_{k=1}^{n} w_k\ds}\right)
	\,.
\end{equation}
In Fig.~\ref{fig:iso-twr} we calculate each sample's $\xi_{\min}\ds$,
then use Eq.~\ref{eq:Hl-angular-scale} to convert this angular
resolution to $\ell_{\max}\ds$: the maximum usable $H_\ell\ds$ (shown
with a dotted line).

Recall that we defined the angular resolution in terms of the smallest
meaningful inter-particle correlation which, according to
Eq.~\ref{eq:Hl-asym}, should mark the beginning of the asymptotic
plateau.  If we look at the two low-multiplicity samples in
Fig.~\ref{fig:iso-twr}, their $\ell_{\max}\ds$ occurs directly after
the start of their respective plateaus --- our definition of
$\xi_{\min}\ds$ is self-consistent.  Note that in the $N^*=126$
sample, a small $\ell=60$ correlation is still visible beyond its
$\ell_{\max}\ds$. This occurs because two particles separated at
$12^\circ$ can also enhance a $6^\circ$ correlation, which makes
$\xi_{\min}\ds$ a \emph{conservative} estimate of the angular
resolution.  Nonetheless, $\xi_{\min}\ds$ correctly identifies when
$H_\ell\ds$ has settled into its asymptotic plateau, and likely
contains no further information (note that in the two high-resolution
samples, the flat plateau is replaced by the angular artifact of the
saturated calorimeter lattice, with repeating overtones as
$\ell\to\infty$).  We have established a link between event
multiplicity $N$ and angular resolution, and used it to define a
conservative estimate for the angular resolution.  In the next
Section, we modify the Fox-Wolfram assumptions and determine how to
use the angular resolution to discard spurious, high-$\ell$
correlations.

\section{Shape functions as low-pass filters}\label{sec:shape-functions}

In order to remove high-$\ell$ correlations that are dominated by
sampling noise, we require a low-pass filter.  Defining a filter is
complicated because each event contains a distinct angular resolution.
Simply cutting off $\ell>\ell_{\max}$ (a step function) would work,
but that scheme gives equal weight to $\ell=2$ and
$\ell=\ell_{\max}$, even though the coarse angular scale of $\ell=2$
makes it \emph{far less} sensitive to small perturbations.\footnote
{Compare the low-$\ell$ components of Figs.~\ref{fig:iso-trk}
  and~\ref{fig:iso-twr}; the calorimeter lattice has almost no effect.
}
Instead we want to use a shape function that emphasizes low~$\ell$
moments and suppresses high~$\ell$ moments. In addition, our function
will have the advantage that experimental resolutions can be
incorporated into the theoretical predictions.

While the discrete event shape $\rho(\vecN{r})$ of
Eq.~\ref{eq:event-shape-delta} was the simplest, most compact way of
summarizing an event's energy distribution from a finite sample, its
$\delta$-distributions imply that each particle's angular position
$\vecN{p}_i\ds$ is known with \emph{infinite} angular resolution.  In
order to accommodate finite angular resolution, we build a low-pass
filter into each particle's angular position --- distributing each
particle in space using a continuous shape function
$h_i\ds(\vecN{r})$:
\begin{equation}\label{eq:rho-shape}
	\rho(\vecN{r}) 
		= \sum_{i=1}^N f_i\ds\,h_i\ds(\vecN{r})
	\,.
\end{equation}
The shape function \emph{smears} each particle about its observed
location $\vecN{p}_i\ds$, preserving its coarse location, but reducing
the angular resolution of $\rho(\vecN{r})$.

\begin{figure}[t]
\centering
\includegraphics[scale=0.5,clip]{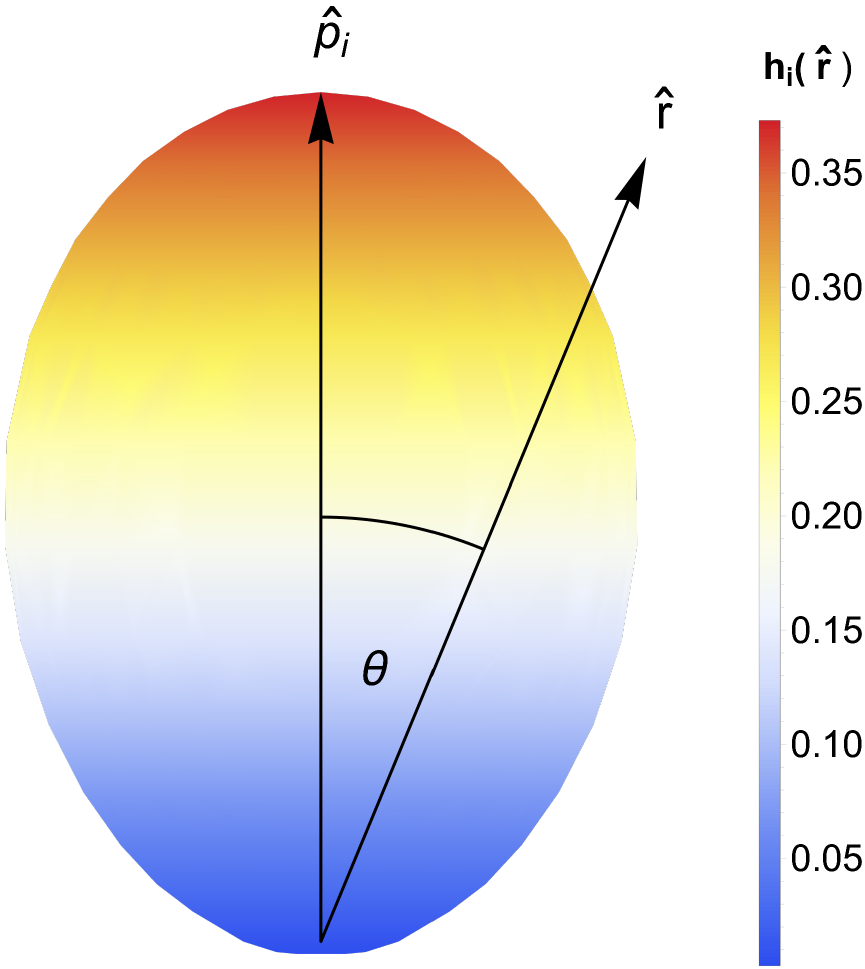}
\caption
{
	A plot of a particle shape function $h_i(\vecN{r})$ that is 
	pseudo-Gaussian in polar angle~$\theta$ 
	(relative to the particle's observed direction of travel $\vecN{p}_i$).
}
\label{fig:Gaussian-shape}
\end{figure}

We are free to choose nearly any shape which is normalizable
($\int_\Omega \diff{\Omega}\,h(\vecN{r})=1$) and acts as a low-pass
filter.  A natural choice is azimuthally symmetric about
$\vecN{p}_i\ds$ and pseudo-normal in polar angle $\theta_i\ds$ (where
$\cos\theta_i\ds\equiv\vecN{r}\cdot\vecN{p}_i$):
\begin{equation}\label{eq:pseudo-normal}
	h_i\ds(\vecN{r})
		= \frac{1}{2\pi\lambda^2(1-e^{-2/\lambda^2})}
		\exp\left(-\frac{(1-\vecN{r}\cdot\vecN{p}_i\ds)}{\lambda^2}\right)
		\overset{\lambda\ll1}{\approx}
		\frac{1}{2\pi\lambda^2}\exp\left(-\frac{\theta_i^2}{2\lambda^2}\right)
	\,,
\end{equation}
This choice has the advantage that it reduces to a Gaussian in the
small angle limit --- a form that maps to the leading Sudakov form
factor in angular-ordered showering.  To relate this distribution's
width $\lambda$ to the sample's angular resolution $\xi_{\min}\ds$, we
solve for the $\lambda$ which encloses a fraction
$u\in\lbrack0,1\rbrack$ of each particle within a circular cap of
angular radius $R$:
\begin{equation}
	\lambda =
	\sin\left(\frac{R}{2}\right)
	\sqrt{\frac{-2}{\log(1-u(1-e^{-2/\lambda^2}))}}
	\,.
\end{equation}
Setting $R = \xi_{\min}\ds$, this transcendental equation
can be solved recursively (starting with $\lambda=0$).\footnote
{For $u\gtrapprox0.9$ and $\xi_{\min}\ds\ll1$, the recursion usually
  stabilizes within machine precision in two iterations (since
  $1-e^{-2/\lambda^2}\approx1$ for $\lambda\ll1$).}

However, by giving particles an extensive shape we can no longer use
the simple linear algebra of Eq.~\ref{eq:H_l-discrete} --- we must
return to original definition:
\begin{equation}
	H_\ell\ds\equiv
		\frac{4\pi}{2\ell+1}
		\sum_{m=-\ell}^{\ell}\abs{\rho_\ell^m}^2	
	\,.
\end{equation}
The $\rho_\ell^m$ integral for a continuous, multi-pronged event shape
is not trivial.  First we use the definition of $\rho(\vecN{r})$ in
Eq.~\ref{eq:rho-shape} to separate the $\rho_\ell^m$~integral into
each particle's shape function
\begin{align}\label{eq:rho-composite}
	\rho_\ell^m 
	& = f_1\ds\,{h_{(1)}\ds}_\ell^m + f_2\ds\,{h_{(2)}\ds}_\ell^m + {\dots} + f_N\ds\,{h_{(N)}\ds}_\ell^m
	\,.
\end{align}
This permits the square modulus $\abs{\rho_\ell^m}^2$ to be expanded
into a series of terms like
${h_{(i)}\ds}_\ell^m{h_{(j)}\ds}_\ell^{m*}$.  Using the $Y_\ell^m$
addition theorem, we can prove that each $H_\ell\ds$ sub-term is
rotationally invariant (like $H_\ell\ds$ as a whole), since $\vecN{r}$
and $\vecN{r}^\prime$ fully factorize, save for the rotationally
invariant $\vecN{r}\cdot\vecN{r}^\prime$;
\begin{align}\label{eq:Hl-sub-term}
	\sum_{m=-\ell}^{\ell}{h_{(i)}\ds}_\ell^m{h_{(j)}\ds}_\ell^{m*}
	& = \sum_{m=-\ell}^{\ell}
		\int_{\Omega}\diff{\Omega}\,
			{Y_\ell^m}^*(\vecN{r})h_{(i)}\ds(\vecN{r})
		\int_{\Omega^\prime}\diff{\Omega^\prime}\,
			Y_\ell^m(\vecN{r}^\prime)h_{(j)}\ds(\vecN{r}^\prime)\\
	& = \int_{\Omega}\diff{\Omega}\int_{\Omega^\prime}\diff{\Omega^\prime}\,
		h_{(i)}\ds(\vecN{r})h_{(j)}\ds(\vecN{r}^\prime)
		P_\ell\ds(\vecN{r}\cdot\vecN{r}^\prime)
	\,.
\end{align}
We can therefore calculate each sub-term in whatever orientation is
the simplest.

\begin{figure}[htb]
\centering
\includegraphics[scale=0.3]{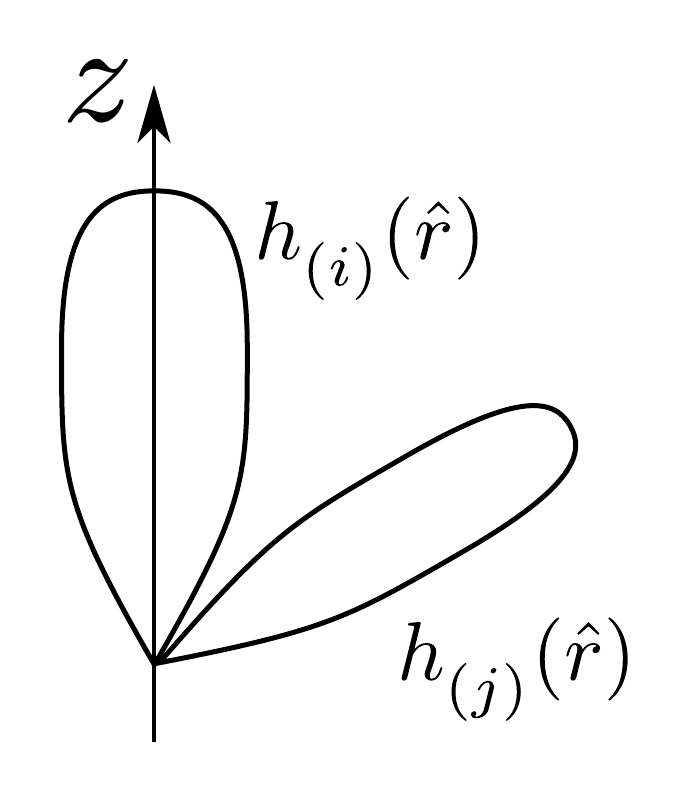}
\caption{
	A pair of shape functions which have been rotated such that 
	$h_{(i)}(\vecN{r})$ is parallel to the $z$-axis (``up''),
	maintaining the interior angle with $h_{(j)}(\vecN{r})$ (``out'').}
\label{fig:up-out}
\end{figure}

A natural choice of orientation is depicted in Fig.~\ref{fig:up-out}:
$h_{(i)}\ds(\vecN{r})$ is oriented ``up'' in the $Y_\ell^m$ system
(parallel to the longitudinal $z$ axis), which tilts
$h_{(j)}\ds(\vecN{r})$ ``out.''  Alternatively, the $j$~shape can be
oriented ``up,'' with $i$ sticking ``out.''  Rotational invariance
makes these choices equivalent, which we can express using an arrow to
denote orientation ($\uparrow$~being ``up'' and $\nearrow$~being
``out'')
\begin{equation}\label{eq:up-out}
	\sum_{m=-\ell}^{\ell}{h_{(i)}^{\;\uparrow}}_\ell^m\, {h_{(j)}^\nearrow}_\ell^m
	=\sum_{m=-\ell}^{\ell}{h_{(i)}^\nearrow}_\ell^m\, {h_{(j)}^{\;\uparrow}}_\ell^m
	\,.
\end{equation}
Now let us constrain $\rho(\vecN{r})$ to only use shape functions
which are azimuthally symmetric about their central axis.  This forces
all ``up'' coefficients to zero except $m=0$, which collapses the
inner products on both sides of the equation to
\begin{equation}\label{eq:up-out-azi-symm}
	{h_{(i)}^{\;\uparrow}}_\ell^0\, {h_{(j)}^\nearrow}_\ell^0
	={h_{(i)}^\nearrow}_\ell^0\, {h_{(j)}^{\;\uparrow}}_\ell^0
	\,.
\end{equation}
Recall that these coefficients are independent scalars from separate
integrals (see Eq.~\ref{eq:Hl-sub-term}); not only will
Eq.~\ref{eq:up-out-azi-symm} hold for \emph{any} azimuthally symmetric
$j$~shape, choosing a particular $j$~shape cannot impose any
constraints on the $i$~shape.

We now choose $h_{(j)}\ds(\vecN{r})=
	\frac{\delta^2(\vecN{r} - \vecN{p}_j\ds)}{\sin\theta_j\ds}$,
a shape whose coefficients are trivial:
${h_{(j)}^{\;\uparrow}}_\ell^0 = 1$ and 
${h_{(j)}^\nearrow}_\ell^0 = P_\ell\ds(\vecN{p}_i\ds\cdot\vecN{p}_j\ds)$.
Plugging these into Eq.~\ref{eq:up-out-azi-symm}, it simplifies to 
\begin{equation}\label{eq:out}
	{h_{(i)}^\nearrow}_\ell^0\, 
	= {h_{(i)}^{\;\uparrow}}_\ell^0\,
	P_\ell\ds(\vecN{p}_i\ds\cdot\vecN{p}_j\ds)
	\,.
\end{equation}
Regardless of the shape of $h_{(i)}\ds(\vecN{r})$,
its ``out'' coefficient can be calculated from its ``up'' coefficient.
Using an overline to signify a scaled, $m=0$ ``up'' coefficient
\begin{align}\label{eq:hl-up}
	{\coeff{h}_{(i)}\ds}_\ell\ds \equiv 
	\sqrt{\frac{4\pi}{2\ell+1}}{h_{(i)}^{\;\uparrow}}_\ell^0
	= \int_\Omega\diff{\Omega}\,P_\ell\ds(\vecN{r}\cdot\vecN{z})\,
	h_{(i)}^{\;\uparrow}(\vecN{r})
	\,,
\end{align}
each sub-term in the $H_\ell\ds$ expansion simplifies to
\begin{equation}
	\frac{4\pi}{2\ell+1}\sum_{m=-\ell}^{\ell}{h_{(i)}\ds}_\ell^m{h_{(j)}\ds}_\ell^{m*}
	= {\coeff{h}_{(i)}\ds}_\ell\ds\, {\coeff{h}_{(j)}\ds}_\ell\ds\,
	P_\ell\ds(\vecN{p}_i\ds\cdot\vecN{p}_j\ds)
	\,.
\end{equation}
Therefore, given an event shape $\rho(\vecN{r})$ composed solely from
particles with azimuthally symmetric shape functions, its power
spectrum is
\begin{equation}\label{eq:Hl-azi-symm}
	H_\ell\ds = 
		(f_i\ds\,{\coeff{h}_{(i)}\ds}_\ell\ds)\,
		P_\ell\ds(\vecN{p}_i\ds\cdot\vecN{p}_j\ds)\,
		(f_j\ds\,{\coeff{h}_{(j)}\ds}_\ell\ds)\,
	\,.
\end{equation}
If every particle uses \emph{the same} shape function,
then this simplifies to an $\ell$-dependent prefactor that
scales the raw, discrete power spectrum of Eq.~\ref{eq:H_l-discrete},
\begin{equation}\label{eq:Hl-azi-symm-same}
	H_\ell\ds = 
	\coeff{h}_\ell^2\times
	\bra{f} P_\ell\ds\big(\ket{\vecN{p}} \cdot\bra{\vecN{p}}\big)\ket{f}
	\,.
\end{equation}

This last scenario is the simplest use-case for shape functions ---
calculate the sample's angular resolution $\xi_{\min}\ds$, then create
\emph{one} pseudo-normal shape that smears each particle by an
appropriate $\lambda$.  According to Eq.~\ref{eq:Hl-azi-symm-same},
plotting $\coeff{h}_\ell^2$ for this shape will reveal its band limit,
and we show several values of $\lambda$ in
Fig.~\ref{fig:Hl-attenuation} (see Appendix~\ref{sec:h_l} for the
detailed calculation of~$\coeff{h}_\ell\ds$).  This plot demonstrates
that the pseudo-normal shape acts as an ideal low-pass filter; it
preserves information at low~$\ell$, and gradually discards
information as $\ell$ increases.  As seen in
Fig.~\ref{fig:Hl-attenuation-log}, $\coeff{h}_\ell\ds$ eventually
enters an approximately exponential decay which removes $H_\ell\ds$'s
asymptotic plateau. This has a meaningful interpretation --- the
discrete event shape created a plateau in $H_\ell\ds$ because
$\delta$-distributions contain every frequency
($\coeff{h}_\ell\ds=1$).  A coarse-graining provided by extensive
shapes imposes a band limit though the asymptotic decay of
their~$\coeff{h}_\ell\ds$.

\begin{figure}[t]
\centering
\subfloat[\label{fig:Hl-attenuation-linear}]{\includegraphics[width=0.5\textwidth]{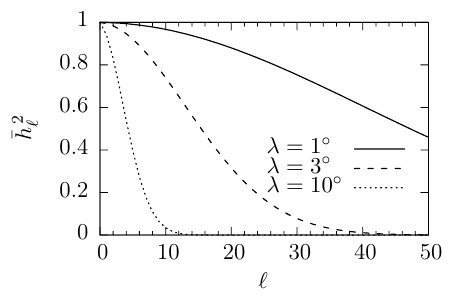}}
\subfloat[\label{fig:Hl-attenuation-log}]{\includegraphics[width=0.5\textwidth]{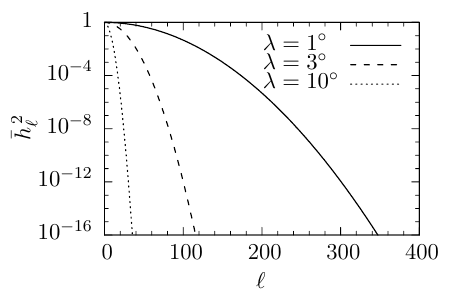}}

\caption{
	The squared ``up'' coefficient $\coeff{h}^2_\ell$ for the 
	pseudo-normal shape function, with several smearing angles $\lambda$
	on a (a) linear and (b) log-linear scale.}
\label{fig:Hl-attenuation}
\end{figure}

\subsection{Shape functions for measurement uncertainty}\label{sec:measurement-error}

Thus far, we have used shape functions to discard small-angle
sampling noise in an under-sampled distribution, so that we can focus
on an its large-angle structure.  However, we can also use shape
functions more traditionally: to encapsulate measurement uncertainty.
For example, if one particular charged track is poorly measured, then
its angular uncertainty may exceed the sample's $\xi_{\min}\ds$, and
its individual shape function should be driven by its measurement
uncertainty.  A more important case is calorimeter towers, whose
intrinsic angular uncertainty \emph{dwarfs} that of tracks, and
usually even $\xi_{\min}\ds$.

A calorimeter is built from multiple layers of cells with varying
segmentation and sensitivity, and non-overlapping centers. This
provides detailed information about the depth and breadth of the
induced particle shower.  ``Towers'' are not bins of energy with clean
edges.  Sophisticated modeling and reconstruction algorithms should be
able to build a probability distribution for the angular position of a
tower's initiating particle, but such modeling is beyond the scope of
this paper.  Instead we consider a pseudo-detector built of towers
that use a uniform \emph{circular} cap of angular radius $R$, which
provides azimuthal symmetry \emph{and} a decent approximation for the
tower (even though the edges are wrong, it creates a low-pass filter
at the correct scale).  We calculate $\coeff{h}_\ell\ds$ coefficients
for a uniform circular cap in Appendix \ref{sec:h_l}.

Consider a track landing near a tower's centroid; while the
track-\emph{centroid} angle
${\xi_{ij}\ds\equiv\arccos(\vecN{p}_i\ds\cdot\vecN{p}_j\ds)}$ is
small, we cannot pinpoint where the energy was deposited.  We should
therefore use the angle between the two objects, which averages over
their constituent shape functions:
\begin{equation}\label{eq:xi_ij-extensive}
	\xi_{ij}^* = \int\diff{\Omega}\int\diff{\Omega^\prime}
	\,h_{(i)}\ds(\vecN{r})\,h_{(j)}\ds(\vecN{r}^\prime)\arccos(\vecN{r}\cdot\vecN{r}^\prime)
	\,.
\end{equation}
Thus, $\xi_{ij}^*$ is used to calculate $\xi_{\min}\ds$ (leading to a
larger value than estimated by Eq.~\ref{eq:angular-resolution}). While
calculating $\xi_{\min}\ds$, tracks should use a shape function based
upon their individual measurement uncertainty. Once the angular
resolution is known, $\xi_{\min}\ds$ sets the minimum smearing for all
objects, and tracks should be smeared with the pseudo-normal shape.
Note that $H_\ell\ds$ already accounts for extensive objects; the
extensive angle (Eq.~\ref{eq:xi_ij-extensive}) is only needed for
$\xi_{\min}\ds$.

\begin{figure}[t]
\includegraphics[width=0.5\textwidth]{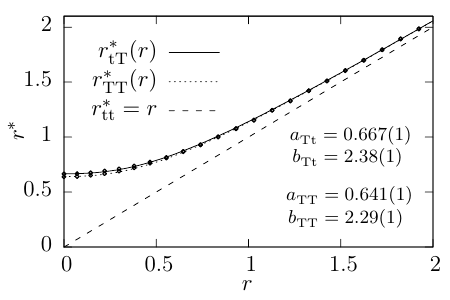}
\caption{The dimensionless angular separation between objects,
  showing the relationship between the extensive angle $r^*$ and the
  raw, inter-centroid angle $r$.  The two are equivalent between
  well-measured tracks.  Approximating towers as uniform circular
  caps, the result of numerical integration ($\circ$) are fit by the
  pseudo-hyperbola (Eq.~\ref{eq:pseudo-hyperbola}) for the
  (t)rack-(T)ower angle ($R_i\ds=0$, $R_j\ds=1^\circ$) and the
  (T)ower-(T)ower angle ($R_i\ds=R_j\ds=1^\circ$).}
\label{fig:rStar}
\end{figure}

In this paper we use a pseudo-detector which assumes exceptional
tracking, such that the difference between $\xi_{ij}^*$
and~$\xi_{ij}\ds$ is dominated by towers.  It is therefore safe to
treat tracks as $\delta$-distributions, so that (t)rack-(t)rack angles
can use $\xi^*_\text{tt}=\xi$.  To calculate the (T)ower-(T)ower
extensive angle~$\xi^*_\text{TT}$, we switch to a relative angular
distance~$r$: the inter-tower angle scaled by the angular radius of
the two-tower system\footnote{ We choose this definition because the
  standard deviations of uncorrelated distributions add in quadrature.
} $R_{ij}\ds = \sqrt{R_i^2 + R_j^2}$;
\begin{align}
	r_{ij} & \equiv \frac{\xi_{ij}\ds}{R_{ij}\ds}\,, & 
	r^*_{ij} & \equiv \frac{\xi^*_{ij}}{R_{ij}\ds}
	\,.
\end{align}
By sending $R_i\ds\to0$, this $r$ also works for the (t)rack-(T)ower
angle $\xi^*_\text{tT}$.

Using the scale-free angle $r$ uncovers a nearly universal mapping
$\xi\mapsto\xi^*$, shown in Fig.~\ref{fig:rStar}. As an object
approaches a tower, the extensive angle $\xi^*$ approaches a minimum
value. Conversely, as the tower gets farther away, it appears
increasingly discrete and $\xi^*\sim\xi$.  The shape of this curve is
well-approximated by the pseudo-hyperbolic function
\begin{equation}\label{eq:pseudo-hyperbola}
	r^*(r) = (a^b + r^b)^\frac{1}{b}
	\,.
\end{equation}
The best-fit parameters accompanying Fig.~\ref{fig:rStar} demonstrate
that track-tower angles $r^*_\text{tT}(r)$ and tower-tower angles
$r^*_\text{TT}(r)$ are barely distinguishable as $r\to0$.  Towers with
non-matching radii ($R_i\ds\neq R_j$) exist on an intermediate curve.
Importantly, these curves are effectively scale-free until the
two-tower radius scale~$R_{ij}\ds$ grows \emph{very} large (about
$30^\circ$), at which point the exponent $b$ becomes larger, causing a
faster decay to the asymptote.  Both fits are systematically biased in
a few places (e.g., a slight overestimation near $r=1$), but recall
that the angular resolution (Eq.~\ref{eq:angular-resolution}) uses the
\emph{geometric} mean to average over scales.  Thus, it is far better
to use $r^*_\text{tT}(r)$ to get a conservative estimate for
$\xi_{ij}^*$, rather than using the raw centroid angle~$\xi_{ij}\ds$,
which would claim that a track striking a tower's center is a
well-measured, small-angle correlation.

\subsection{Infrared and collinear safety}\label{sec:IRC}

In this section, we have taken the QCD radiation spectrum of Fox and
Wolfram and expanded it with a comprehensive framework of shape
functions to account for sampling granularity and measurement
uncertainty.  Putting these pieces together for the first time, we
immediately see that shape functions are essential for keeping
$H_\ell\ds$ infrared and collinear safe. This is the final piece we
need to use the QCD radiation spectrum.

Let us return to the two events of Fig.~\ref{fig:Hl-intro}, which were
analyzed via a truth-level detection of their measurable final-state
particles.  We now filter those particles through our pseudo-detector
(using $\Omega_{\twr}\ds=(6^\circ)^2$, and cuts
$\eta_{\max}^{\trk}=\eta_{\max}^{\twr}=3$, and
$p_T^{\min}=\unit[300]{MeV}$), calculate the angular resolution
$\xi_{\min}\ds$ using extensive inter-particle angles~$\xi^*_{ij}$,
then smear the tracks via a pseudo-normal shape function to calculate
$H_\ell\ds$ (solving $\lambda$ for $u=90\%$ of the track shape within
in a cap of radius $R=\xi_{\min}\ds$).  This scheme produces a smooth
event shape~$\rho(\vecN{r})$ with a sample's ``natural resolution.''
Yet examining its power spectrum by eye does not provide much insight
--- it looks like the moments in Fig.~\ref{fig:Hl-intro} multiplied by
a low-pass filter similar to Fig.~\ref{fig:Hl-attenuation}.  Luckily,
there is a much more physical way to view the power spectrum.

In their seminal papers \cite{Fox:1978vu,Fox:1978vw}, Fox and Wolfram
introduced an ``autocorrelation'' function which we relabel the
``angular correlation function;'' it uses $H_\ell\ds$ as the
coefficients in a Legendre series for inter-particle angle $\xi$:
\begin{equation}
	F(\cos\xi)=\sum_{\ell=0}^\infty (2\ell+1)H_\ell\ds\,P_\ell\ds(\cos\xi)
	\,.
\end{equation}
The area under the angular correlation function is always two (since
$\int_{-1}^1\diff{z}\,P_\ell\ds(z)=2\,\delta_\ell^0$ and $H_0\ds=1$).
A peak at $\cos\xi$ indicates that energy within the sample is
correlated at that angle, with the area under the peak equal to the
collective weight $w=4\sum f_i\ds f_j\ds$ of all pairs separated by
$\xi$ (relative to each other, not to any particular axis).  For
example, events containing prongs of collinear energy should have a
large $F$ at $\cos\xi=1$.

\begin{figure}[t]
\subfloat[\label{fig:2-jet-angular}2-jet-like]{
\includegraphics[width=0.5\textwidth]{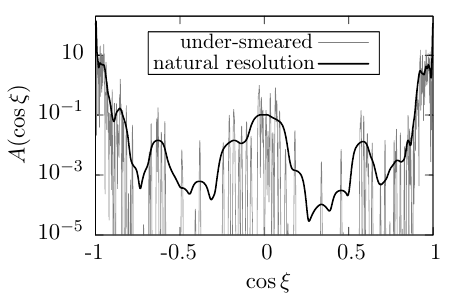}}
\subfloat[\label{fig:3-jet-angular}3-jet-like]{
\includegraphics[width=0.5\textwidth]{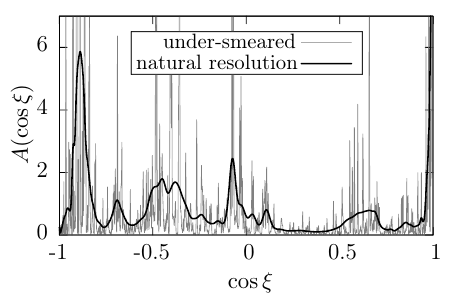}}
\caption{The angular correlation function for two 
$e^+e^-\to q\bar{q}g$ events at $\sqrt{S}=\unit[250]{GeV}$ which are
(a)~2-jet-like and (b)~3-jet-like.}
\label{fig:angular-natural-under}
\end{figure}

Clearly, if $H_\ell$ flattens to an asymptotic plateau, $F$ will never
converge for finite $l$, so it is not actually possible to plot $F$
from the Fox-Wolfram event shape.  However, we can create a nearly
discrete event shape by smearing every detected track or tower with a
pseudo-Gaussian \emph{ten times} thinner than the one prescribed by
the sample's angular resolution~$\xi_{\min}\ds$.  In
Figure~\ref{fig:angular-natural-under}, we compare this
``under-smeared''~$\rho(\vecN{r})$ to the natural resolution shape.
The difference is quite striking.  For both events, the under-smeared
$F$ contains extremely narrow correlations from individual particle
pairs. Conversely, the appropriate amount of smearing gathers these
ultra-fine details into a smooth peak.  This permits us to verify the
event's overall shape: the 2-jet-like event
(Fig.~\ref{fig:2-jet-angular}, shown on a log-scale) is extremely
back-to-back, since the largest correlation other than $\cos\xi=\pm1$
is incredibly small (i.e., $F(\cos90^\circ)\approx0.1$).  This is much
different than the 3-jet-like event (Fig.~\ref{fig:3-jet-angular}),
which has three large peaks at $\cos\xi<0$, corresponding to the three
inter-jet angles (c.f.\ the inset of Fig.~\ref{fig:3-jet-unsmeared}).

It is clear, given the weight ${w_{ij}\ds=f_i\ds f_j\ds}$ of each
correlation, that $H_\ell\ds$ (and thus $F$) is intrinsically
insensitive to infrared radiation (which has $f\ll1$ by definition).
Figure~\ref{fig:angular-natural-under} now shows that shape functions
make $H_\ell\ds$ insensitive to collinear radiation (the fine
structure visible in the under-smeared $F$).  This fine structure is
only probed by high-$\ell$ moments (Eq.~\ref{eq:Hl-angular-scale}), a
relationship we can make explicit by examining a particle splitting to
two nearly parallel particles ($a\to b\,c$).  First we examine $a$'s
total contribution to the power spectrum \emph{before} the splitting
(where $z_{aj}\ds\equiv\hat{p}_a\ds \cdot \hat{p}_j\ds$)
\begin{equation}
	H_{\ell,\,a}\ds = f_a\ds \sum_{j}\ds f_j\ds\, P_\ell\ds(z_{aj}\ds)
	\,.
\end{equation}
When particle $a$ splits, $\hat{p}_{b/c}\ds = \hat{p}_a\ds +
\vec{\delta}_{b/c}$ (for tiny $\vec{\delta}_{b/c}$), this contribution
becomes
\begin{align}
	H_{\ell,\,a}\ds
	& = f_b\ds \sum_{j}\ds f_j\ds P_\ell\ds(z_{aj}\ds + \delta z_{bj}\ds)\nonumber
	+ f_c\ds \sum_{j}\ds f_j\ds P_\ell\ds(z_{aj}\ds + \delta z_{cj}\ds)
	\,.
\end{align}
Because ${f_a\ds = f_b\ds + f_c\ds}$, the total contribution
$H_{\ell,\,a}\ds$ is perturbed only in the $P_\ell\ds$ terms.  And
only when $\ell$ becomes large --- so that $P_\ell\ds(z)$ is highly
oscillatory at the scale of $\delta z$ --- can a small $\delta z$ give
rise to significant changes in~$H_\ell\ds$.  Hence, a low-pass filter
tailored to an event's information content is essential for keeping
the power spectrum infrared and collinear safe.  With this we now have
a complete framework in which to encode the full energy and angular
correlated information of a QCD event.


\section{Conclusions}\label{sec:conclusions}

In this paper we address some deficiencies of the energy-weighted QCD
angular power spectrum and its artifacts due to finite sampling.  We
extend the power spectrum introduced by Fox and Wolfram
\cite{Fox:1978vu,Fox:1978vw} by replacing an energy density distribution
constructed out of measured point-like particles with a shape function
that acts as a band pass filter to suppress these artifacts.  The
shape function naturally accommodates the measurement resolutions of
tracks and calorimeter cells, and other experimental complications,
such as false peaks arising from holes in the detector.  While our
shape function maps directly to leading QCD radiation at small angles,
other functions that preserve the smallest measurable angle constraint
could be used to more adequately map out detector-dependent shapes
unique to a particular design.

By construction, the moments $H_\ell\ds$ of the power spectrum are
infrared safe.  Our use of shape functions to smear an event to its
sampled resolution guarantees collinear safety by discarding
correlations at angular scales smaller than the information content of
the event.  Hence, we have established a framework in which
information encoded in the power spectrum can be used to extract
well-defined observables using the fully-correlated data from an
entire QCD event.

The next step is to build studies to explore QCD phenomena at all
scales. An initial foray may be found in
Refs.\ \cite{Sullivan:2019kqy,Pedersen24}, where jet-like physics in
$e^+e^-$ events is extracted with higher precision than that provided
by current sequential jet algorithms in the presence of pileup.  We
should also improve theoretical predictions by calculating
next-to-leading order corrections to the QCD power spectrum
itself. From there we can begin to map the effects of jet substructure
--- noting that we will capture the effects of radiation emitted at
angles larger than normally captured by sequential jet definitions.

This paper focuses on $e^+e^-$ events and calculations in the center
of momentum frame which can be derived analytically. While it is
straightforward to compute the power spectrum for hadronic collisions,
it can only be performed numerically. The next step is to provide the
boosted calculations necessary to compare numerical simulations of the
power spectrum to QCD jet data provided by the ATLAS and CMS
Collaborations at the CERN Large Hadron Collider.  Finally, if a power
spectrum can be calculated using the fundamental QCD degrees of
freedom found in heavy-ion collisions, this framework can be used to
understand short- and long-range correlations in heavy ion data on an
event-by-event basis.


\begin{acknowledgments}
This paper is based upon work supported by the U.S. Department of
Energy, Office of Science, Office of High Energy Physics under Award
Number DE-SC-0008347.
\end{acknowledgments}

\appendix
\section{Connecting $\bs{\braket{f}{f}}$ to particle multiplicity $\bs{N}$}
\label{sec:ff-N}

In Section~\ref{sec:power-spectrum} we explore the consequences of the
connection between $\braket{f}{f}$ and particle multiplicity. In this
appendix we present the derivation of the relationship (taken from
Ref.\ \cite{Pedersen2018}).

\subsection{Why $\bs{\braket{f}{f} \propto N^{-1}}$}%

To study the expected value $\Ex{\braket{f}{f}}$, we assume that 
there is a generic physics process where 
particle energy follows some smooth probability distribution $h(E)$, 
but where particle multiplicity~$N$ is variable.
Converting this $h(E)$ distribution to energy fraction $h(f)$, 
we find the undesirable property that the mode
scales with~$N$ (since the mean $\Ex{f} = N^{-1}$ by construction).
It is therefore useful to define the scale-free energy fraction
\begin{equation}	
	\fv \equiv \frac{f}{\Ex{f}} = N f
	\,.
\end{equation}
We then require that $h(\fv)$ has a finite variance
\begin{equation}
	\text{Var}(\fv) = \Exx{\fv^2}-\Exx{\fv}^2 = \Exx{\fv^2}-1
	\,.
\end{equation}

We now construct an energy fraction vector $\ket{f}$ for 
a random instance of this physics process;
we draw $N$ energy fractions $\fv$ from $h(\fv)$, then normalize to their sum:
\begin{equation}\label{eq:fv}
	\ket{f} = \frac{\lbrace \fv_1\ds, \fv_2\ds, \dots{}, \fv_N\ds \rbrace}
	{\fv_1\ds + \fv_2\ds + \dots{} + \fv_N\ds}\,.
\end{equation}
The expectation value of $\braket{f}{f}$ is therefore
\begin{equation}\label{eq:f2-epect-orig}
	\Ex{\braket{f}{f}}= 
	\int\diff \fv_1\ds\, \diff \fv_2\ds \dots{} \diff \fv_N\ds
	\frac{\fv_1^2 + \fv_2^2 + \dots{} + \fv_N^2}
	{(\fv_1\ds + \fv_2\ds + \dots{} + \fv_N\ds)^2}
	\,h(\fv_1\ds)\,h(\fv_2\ds)\dots{}h(\fv_N\ds)
	\,.
\end{equation}
Provided that each $\fv_i\ds$ is independent (i.e., all correlations
are built into the shape of $h(\fv)$), we can treat each $\fv_i\ds$
separately and use the linearity of expectation to obtain
\begin{equation}\label{eq:f2-expect}
	\Ex{\braket{f}{f}}
		= \frac{N\,\Exx{\fv^2}}{(N\,\Exx{\fv})^2}
		= \frac{1}{N}\,\Exx{\fv^2}
		= \frac{1}{N}(1 + \text{Var}(\fv))
		\,.
\end{equation}
Thus, we expect the height of the power spectrum's asymptotic plateau
${H_\ell\ds\sim\braket{f}{f}}$ to be inversely proportional to
particle multiplicity, but somewhat larger than~$1/N$ since
${\text{Var}(\fv)\ge0}$.

\subsection{The smallest possible $\bs{\braket{f}{f}_{\min}\ds = N^{-1}}$}
\label{appx:H_l-min}

As a check of Eq.~\ref{eq:f2-expect}, we calculate the smallest
possible value of $\braket{f}{f}$ \emph{without} assuming some
well-behaved physics process with a limiting distribution $h(\fv)$.
We define a normalized energy fraction vector for some arbitrary set
of particles:
\begin{equation}
	\ket{f} = \left\lbrace f_1\ds,\, \dots{},\, f_{N-1}\ds,\, f_{N}\ds\right\rbrace,
	\qquad\text{where}\quad f_N\ds \equiv 
	\Big(1 - \sum_{i=1}^{N-1}f_i\ds\Big)\ge 0\,.
\end{equation}
Evaluating the gradient $\vec{\nabla}\braket{f}{f}$, 
each energy fraction $f_i\ds$ minimizes $\braket{f}{f}$ when
\begin{equation}
	\partial_i\ds \braket{f}{f} 
	= \partial_i\ds (f_i^2 + f_N^2)
	= 2 (f_i\ds - f_N\ds) = 0
	\,.
\end{equation}
Thus, $\braket{f}{f}$ is minimized when $f_i\ds = f_N\ds$, 
and if every energy fraction is equal to the final energy fraction, 
then all $f_i\ds$ must be the same: $f_i\ds = N^{-1}$.
Therefore,
\begin{equation}
	\braket{f}{f}_{\min}\ds = N(N^{-2}) = \frac{1}{N}\,.
\end{equation}
This result corresponds exactly to $\text{Var}(\fv) = 0$ for
Equation~\ref{eq:f2-expect}.

\section{Calculating a shape function's ``up'' coefficient $\bs{\coeff{h}_\ell\ds}$}
\label{sec:h_l}

The power spectrum $H_\ell\ds$ for an event shape $\rho(\vecN{r})$
built from azimuthally symmetric shape functions can be calculated
using Eq.~\ref{eq:Hl-azi-symm} --- provided one knows the ``up''
coefficients~$\coeff{h}_\ell\ds$ (Eq.~\ref{eq:hl-up}).  In this
appendix we follow Ref.\ \cite{Pedersen2018} to calculate
$\coeff{h}_\ell\ds$ for the shape functions used in this paper.  For
each calculation, rotating the shape's centroid ``up''
($\vecN{p}_i\ds\mapsto\vecN{z}$) permits the useful change of variable
${z\equiv\cos(\theta)=\vecN{r}\cdot\vecN{z}}$.  Normalization
constants will appear as $C$.  Note that the $\coeff{h}_0\ds$ integral
is simply the normalization condition, so $\coeff{h}_0\ds=1$ for every
shape function.

\subsection{The pseudo-normal shape function}
\label{sec:hl-pseudo-normal}

The ``up'' coefficient (Eq.~\ref{eq:hl-up}) for 
the pseudo-normal shape function (Eq.~\ref{eq:pseudo-normal}) is
\begin{align}
	\coeff{h}_\ell\ds
	= \int_0^{2\pi}\diff{\phi}\int_{-1}^{1}\diff{z}\,P_\ell\ds(z)\,
		C\,e^{-(1-z)/\lambda^2}
	=2\pi \,C \int_{-1}^{1}\diff{z}\,P_\ell\ds(z)\,
	e^{-(1-z)/\lambda^2}
	\,.
\end{align}
To compute this integral, we can use a Legendre identity
\begin{equation}\label{eq:Pl_to_deriv}
	(2\ell+1)P_\ell\ds(z) = \frac{\diff{}}{\diff{z}}
	\left\lbrack
		P_{\ell+1}\ds(z) - P_{\ell-1}\ds(z)
	\right\rbrack
	\,.
\end{equation}
This allows us to define an integral
$A_\ell\ds\equiv\coeff{h}_\ell\ds/(2\pi\,C)$, and set it up for
integration by parts:
\begin{equation}\label{eq:int-by-parts}
	A_\ell\ds = \int_{-1}^1 \diff{z}\,e^{-(1-z)/\lambda^2}P_\ell\ds(z)
	= \int_{-1}^1 \diff{z}\,e^{-(1-z)/\lambda^2}
	\frac{1}{2\ell+1}
	\frac{\diff{}}{\diff{z}}
	\left\lbrack
		P_{\ell+1}\ds(z) - P_{\ell-1}\ds(z)
	\right\rbrack
	\,.
\end{equation}
The boundary term vanishes because $\lbrack
P_{\ell+1}\ds(\pm1)-P_{\ell-1}\ds(\pm1)\rbrack=0$.  The surviving term
is
\begin{equation}
	A_\ell\ds = \frac{1}{2\ell+1}
	\int_{-1}^1 \diff{z}\,\frac{e^{-(1-z)/\lambda^2}}{\lambda^2}
	(P_{\ell-1}\ds(z) - P_{\ell+1}\ds(z))
	= \frac{1}{\lambda^2(2\ell+1)}(A_{\ell-1}\ds-A_{\ell+1}\ds)
	\,.
\end{equation}
Since $\coeff{h}_\ell\ds\propto A_\ell\ds$, this result can be 
rearranged into a recurrence relation for $\coeff{h}_\ell\ds$:
\begin{equation}\label{eq:hl-pseudo-normal}
	\coeff{h}_{\ell+1}\ds = -(2\ell+1)\lambda^2\, \coeff{h}_{\ell}\ds + \coeff{h}_{\ell-1}\ds
	\,.
\end{equation}
We initialize this recursion with $\coeff{h}_0\ds=1$ (i.e.,
$h(\vecN{r})$ is normalized) and
\begin{equation}
	\coeff{h}_1\ds = \frac{1}{\tanh(\lambda^{-2})}-\lambda^2
	\,.
\end{equation}
For a demonstration of how to correct the recursion's numerical
instability as $\coeff{h}_\ell\ds\to0$ see Ref.~\cite{Pedersen2018} .

\subsection{A circular cap}

\newcommand{\OmegaT}{\Omega_\twr\ds}

To approximate calorimeter towers as uniform circular caps of angular
radius $R$ and solid angle~$\OmegaT=2\pi\,(1-\cos R)$, we integrate
upwards from $z=\cos R=1-a$ (where $a\equiv\frac{\OmegaT}{2\pi}$):
\begin{equation}
	\coeff{h}_\ell\ds
	= \frac{1}{\OmegaT}\int_0^{2\pi}\diff{\phi}
	\int_{1-a}^{1}\diff{z}\,P_\ell\ds(z)
	= \frac{1}{a\,(2\ell+1)} \lbrack P_{\ell-1}\ds(1-a) - P_{\ell+1}\ds(1-a)\rbrack
	\,.
\end{equation}
This uses Eq.~\ref{eq:Pl_to_deriv} to integrate $P_\ell\ds(z)$, and
relies on the identity $\lbrack P_{\ell+1}\ds(\pm1)-P_{\ell-1}\ds(\pm1)\rbrack=0$.

\bibliography{PowerSpectrum}

\end{document}